\documentclass{aastex62}
\begin{document}

\title{Gender and the Career Outcomes of PhD Astronomers in the United States}
\email{d.a.perley@ljmu.ac.uk}

\author[0000-0001-8472-1996]{Daniel A. Perley}
\affil{Astrophysics Research Institute, Liverpool John Moores University, \\ IC2, Liverpool Science Park, 146 Brownlow Hill, Liverpool L3 5RF, UK}

\def \Nall {1154} 
\def \Nprog {28} 
\def \Nnrc {21} 
\def \Naip {24} 
\def \Nallyr {88} 
\def \Nomit {91} 
\def \Nunkgender {36}
\def \Nsamp {1063} 
\def \Nsampm {748} 
\def \Nsampf {315} 
\def \pctm {70.4} 
\def \pctf {29.6} 
\def \pctaip {70} 
\def \Nastro {672}
\def \Nleft {273}
\def \Npostdoc {118}
\def \mpctjob {71}  
\def \fpctjob {71}  
\def \mpctleave {29}  
\def \mpctleaveu {4}  
\def \mpctleavel {3}  
\def \fpctleave {29}  
\def \fpctleaveu {6}  
\def \fpctleavel {5}  
\def \astrop {0.87} 
\def \pchisqe {0.04}
\def \pchisqj {0.17}

\def \Nomityr {37} 
\def \Nyr {1026}    
\def \mmean {4.86} 
\def \mmeanunc {0.12}
\def \fmean {4.41} 
\def \fmeanunc {0.16}
\def \gap {0.45} 
\def \gapunc {0.40} 
\def \gapp {0.045} 
\def \pFlaherty {0.0001} 
\def \pcthiretwelve {65}    
\def \pcthiretwelvem {65}   %
\def \pcthiretwelvef {66}   %
\def \pctleavetwelve {27}   %
\def \pctleavetwelvef {27}  %
\def \pctleavetwelvem {27}  %
\def \pcttemptwelve {8}  %
\def \Hfm {1.08} 
\def \Hfmu {0.20} 
\def \Hfml {0.17} 
\def \Hfmp {0.38} 
\def \Lfm {1.03} 
\def \Lfmu {0.31} 
\def \Lfml {0.24} 
\def \Lfmp {0.81} 

\begin{abstract}
I analyze the postdoctoral career tracks of a nearly-complete sample of astronomers from \Nprog\ United States graduate astronomy and astrophysics programs spanning 13 graduating years ($N=\Nsamp$).  A majority of both men and women (\pcthiretwelvem\% and \pcthiretwelvef\%, respectively) find long-term employment in astronomy or closely-related academic disciplines.  No significant difference is observed in the rates at which men and women are hired into these jobs following their PhDs or in the rates at which they leave the field.  Applying a two-outcome survival analysis model to the entire data set, the relative academic hiring probability ratio for women vs.\ men at a common year post-PhD is  $H_{F/M} = \Hfm^{+\Hfmu}_{-\Hfml}$; the relative leaving probability ratio is $L_{F/M} = \Lfm^{+\Lfmu}_{-\Lfml}$ (95\% CI).  These are both consistent with equal outcomes for both genders ($H_{F/M} = L_{F/M} = 1$) and rule out more than minor gender differences in hiring or in the decision to abandon an academic career.  They suggest that despite discrimination and adversity, women scientists are successful at managing the transition between PhD, postdoctoral, and faculty/staff positions.
\end{abstract}

\section{Introduction}

Women face a variety of obstacles in the academic workplace, particularly in fields such as astrophysics in which they are historically a minority.   Despite strides in recent decades, harassment, unconscious bias, and demands on time (e.g., need for female representation on committees in departments with few women) all fall more heavily on women than on men \citep{MossRacusin+2012,Sleeth+2017,NA2018}.  Women may also face a variety of social pressures more acutely than men outside the workplace, such as a stronger expectation to have children and to be the primary caregiver \citep{Cech+2019}.

The potential effects of gender discrimination on early-career scientists (including PhD students and postdocs) have drawn special attention.  Junior scientists are particularly vulnerable to the effects of institutional biases, due to a lack of long-term job security and dependence on a supervisor or other senior figure for their future career development.  The demands of academic career advancement (which often involves holding short-term postdoctoral positions and numerous relocations) also clash directly with non-academic pursuits, such as a desire to start a family, in a way that can be particularly acute during this period.  Two-body hiring issues further amplify these concerns.

While the existence of these factors is well-documented, there has been little published work on their impact on the careers of scientists in practice.  While it is reasonable to assume that the additional challenges faced by women are a leading cause of their lower representation in the physical sciences (including astronomy, where the fraction of women is approximately 15\% for senior positions and 30\% for early-career positions; \citealt{Hughes+2014}), this can be difficult to show in practice, given the complexities of the lives of individuals and of real-world academic hiring.

Past research on this issue has generally relied on cross-sectional snapshot studies, or on surveys of individuals' past experiences. These types of studies are limited by survivorship bias (individuals who left the field are generally not represented) and have drawn conflicting conclusions as to whether the pipeline between graduate school and a STEM career is leakier for women than it is for men. \citep{Hoffman+2004,Bagenal+2004,Ivie+2005,NRC2010}.

The ideal tool to investigate these effects would be a large-scale longitudinal study in which many hundreds of individuals were tracked starting early in their careers and continuing until they secured long-term employment within their discipline or until they left to pursue another career.  While worthwhile, such an effort would be slow---requiring years if not decades of monitoring and (likely) reliance on self-reporting of the individuals being studied.  As a result there are few studies of this type, most of which have been limited to a relatively short time period (e.g. the study of \citealt{Ivie+2016}, which was effectively restricted to the PhD-to-postdoc transition).  Also, few such studies have been devoted specifically to astronomers (or even to physicists more generally), even though large differences between fields in career-transition gender disparity have been reported \citep{Shaumann2017}.

Fortunately, in this digitally interconnected era, it is no longer necessary to rely on individuals themselves to self-report data.  PhD alumni and dissertation lists are available online, and it is routine for young professionals (both in and out of research careers) to post their CV data publicly on the internet as well---enabling construction of an instant de-facto longitudinal study using only public information.

Our study is inspired by the recent arXiv posting of Flaherty (2018; hereafter F18), who studied the PhD-to-faculty times of astronomers using a sample collected from a public rumor mill website.  Like them, we study astronomers and focus on the phase between PhD and starting a permanent career (that is, the postdoctoral phase, which we also take to include adjunct, lecturer, and short-contract, soft-money positions).  However, unlike them, we monitor the career tracks of PhD recipients \emph{regardless of outcome}, allowing us to draw conclusions about the relative proportions staying in or leaving academia (and the times at which they were hired or left) directly.  We also employ a formal non-parametric statistical analysis and do not rely on a tuned-by-eye labor market model, nor do we rely on the incomplete (and potentially biased) sampling of a rumor mill website.

\section{Data}

The sample is drawn from public PhD alumni and dissertation lists posted on the webpages of major PhD-granting graduate programs across the United States.  We attempted to find all such listings by searching the webpages of 34 medium-to-large US PhD programs in astrophysics\footnote{We exclude programs in planetary science and programs with $<10$ total PhD students reported between 2000--2012.} as listed in the American Institute of Physics (AIP) roster of astronomy programs\footnote{https://www.aip.org/statistics/rosters/astronomy}.  Only programs which provided complete lists (not ``selected'' alumni) were used.  We were able to find \Naip\ such listings: Arizona, UC Berkeley, UC Los Angeles, Caltech, Chicago, Florida, Georgia State, Harvard, Hawaii, Illinois, Maryland, Massachusetts, Michigan, Michigan State, New Mexico State, Ohio State, Princeton, Penn State, Rice, Virginia, Washington, Wisconsin, Wyoming, and Yale.\footnote{Medium-to-large astronomy programs not represented in the sample are Boston, UC Santa Cruz, Colorado, Cornell, Columbia, Indiana, Minnesota, MIT, Johns Hopkins, and Texas.  Small or defunct programs ($<10$ graduates in the AIP roster) are BYU, Case Western, Florida Tech, Iowa State, Pittsburgh, and Tufts.}  Additionally, we searched the websites of a number of physics programs with significant astrophysics components not listed in the AIP astronomy roster and found complete alumni lists for four additional programs: Alabama, Clemson, Dartmouth, and Rochester.  Our final list of \Nprog\ programs is reasonably representative of US astronomy PhD programs in most respects (e.g., geography, prestige of program, scientific focus of department) although it will under-represent astronomers graduating from predominantly physics programs.

We downloaded all names and PhD years from these lists into a spreadsheet (in the case of joint astronomy and physics departments, non-astronomy PhD theses were excluded).  We restricted this sample to the 13-year period between 2000--2012 (inclusive), producing an initial sample of \Nall\ PhDs (about \Nallyr\ per year).  This sample includes \pctaip\% of the PhDs awarded in AIP-listed astronomy programs during this period (and roughly 35\% of all US astrophysics-related PhDs; \citealt{Metcalfe+2008}).

Gender was recorded (as M or F) for each individual on the basis of their first name where possible.  In cases where this was ambiguous, we used an online search engine to find images of the individual or articles referring to them with third-person gendered pronouns\footnote{We recognize that the gender binary is incomplete and does not capture a wide range of individuals who do not fall within these two categories, as is the practice of assigning a gender based only on names or images.  For this work we make the assumption that the number of non-binary or otherwise incorrectly-gendered individuals is not large enough to affect our calculations.}.

Career paths (specifically, PhD year and the date and location of the first long-term appointment) were determined from online CV's, university profiles, from social media sites, from other web sources such as news articles, or from paper affiliations.  When the date could not be inferred exactly (e.g. when inferring from paper affiliations in the presence of a publication gap) we took the average of the last-available pre-hire record and first-available post-hire record.

For a small number of individuals, no recent information on their career status could be discerned: there were no websites, articles, or scientific papers associated with them in many years and it could not be determined what their current location was, although it was clear that they were research-active in the past.  Generally, they were presumed to have left the field following the date of their most recent paper.  However, if there was any evidence that they had shifted into a non-research track but remained within astrophysics, or if the lack of information originated because their name was very common or foreign and ambiguously transliterated into Western writing (making search engines or ADS unreliable), they were omitted instead.  This omission may produce a slight bias (PhD recipients who left astrophysics or moved abroad are more likely to be untraceable).  However, we do not expect this to be gender-dependent, and less than $10$\% of the initial sample is affected affected by omissions for this reason so its effect in practice will be small.

It was not always straightforward to determine whether a position was temporary or long-term.  Many job titles (``associate researcher'', ``research scientist'', ``research professor'') could refer either to career scientists or to late-term postdocs or soft-money researchers on short contracts, and outside traditional university environments the distinction between long-term and short-term positions is not a sharp one.  Where possible we looked up the job description on the employer's website to discern whether it was an independent position with the expectation of lasting many years (even without formal tenure), or if it was contract-based and associated with a PI or lab.  If the nature of the role could not be determined and the job title was ambiguous, the job was assumed to be long-term.\footnote{To some extent this choice was arbitrary: many such ambiguous jobs are likely to be soft-money or grant-supported hires without long-term security.  However, in practice most such positions did last for many years and exceedingly few individuals moved out of astrophysics afterwards, suggesting that it is reasonable to treat them as long-term employment.}

In a few cases it was difficult to define whether the individual was working within astrophysics or not, despite knowledge of their place of employment.  Some were working in universities but in departments outside astronomy or physics, or in non-astrophysics branches of NASA.  Others were employed in private industry, but working in areas with some connection to astrophysics (e.g. aerospace, as a contractor for a NASA mission, in public science policy), or were teaching physics in a high school.  These were generally treated as having left the field, unless the individual appeared to retain a direct connection with astrophysics research or higher education.

A few individuals left to pursue another degree; we record their departure date from astrophysics as the year they began their subsequent studies but classify their career as the category they eventually became employed in.

We excluded cases where we were unable to determine information critical for the analysis: in particular if we could not determine the graduate's gender, or any meaningful information with which to determine the nature of their job.  A small number of individuals who passed away while postdocs, or who were mature students at the time of their PhD and subsequently retired, were also excluded.  Individuals whose career path could be determined but no useful constraint on the hiring date (within $\pm$1 year) was available were excluded from time-based survival analysis calculations but not from general outcome statistics.

\begin{table}[t]
\begin{center}
\begin{tabular}{|l|r|r|l|}
\hline			
\textbf{Outcome by Category} & F  & M  & \%F \\
\hline
Professor (R1, tenure-track) & 41 & 111 & 27 $\pm$ 7 \\
Professor (all other)        & 81 & 155 & 34 $\pm$ 6 \\
Staff scientist / technician & 73 & 211 & 26 $\pm$ 5 \\
\hline
Non-astrophysics             & 81 & 192 & 30 $\pm$ 6 \\
\hline
\hline
Still postdoc / adjunct      & 39 & 79  & 33 $\pm$ 9 \\
\hline
Omitted\footnote{An additional 36 were removed because their genders are unknown.}    & 15 & 40  & 27 $\pm$ 12 \\
\hline
\end{tabular}
\quad
\begin{tabular}{| l | r | r | l |}
  \hline			
  \textbf{Outcome by Employer} & F  & M & \%F \\
  \hline
  \textbf{Astrophysics:}       & \textbf{195} & \textbf{477} & \textbf{29 $\pm$ 4} \\
  University (R1)              & 46 & 141 & 25 $\pm$ 6 \\
  University (R2)              & 15 & 22  & 41 $\pm$ 17 \\
  University (R3/M)            & 21 & 36  & 37 $\pm$ 13 \\
  University (foreign)         & 10 & 34  & 23 $\pm$ 13 \\
  Small college                & 25 & 30  & 45 $\pm$ 14 \\
  Observatory / NASA / lab     & 27 & 78  & 26 $\pm$ 9 \\ 
  Other astrophysics (US)      & 27 & 82  & 25 $\pm$ 8 \\
  Other astrophysics (foreign) & 24 & 54  & 31 $\pm$ 11 \\
  \hline
  \textbf{Non-astrophysics:}   & \textbf{81} & \textbf{192} & \textbf{30 $\pm$ 6} \\
  Univ. or NASA, not astro     & 5  & 4   & 56 $\pm$ 33 \\
  High school / education      & 6  & 8   & 43 $\pm$ 27 \\
  Government / military        & 6  & 13  & 32 $\pm$ 33 \\
  Private industry             & 53 & 149 & 26 $\pm$ 6 \\
  Unknown                      & 11 & 18  & 38 $\pm$ 19 \\
  \hline
\end{tabular}
\caption{Career outcomes (defined as the first long-term job) for men and women astronomy PhDs in this study, showing absolute counts for each gender and the percentages of women (with half-width of the 95\% binomial confidence interval).   Left: simplified outcomes by job title and employer.  Right: detailed outcomes by employer.  Note that both tenure-track and non-tenure-track jobs are included in the ``R1'' row in the table at right.}
\end{center}
\end{table}

Out of the initial sample of \Nall, we removed \Nomit\ individuals for the various reasons described above, leading to a final sample of \Nsamp\ for the outcome analysis (a further \Nomityr\ individuals were excluded from hiring-time based analyses only).  Of these, \Nsampm\ are male (\pctm\%) and \Nsampf\ (\pctf\%) are female, consistent with statistics on the gender ratio of astronomy PhDs compiled elsewhere (\citealt{Hughes+2014}).  Within this sample, \Nastro\ progressed to long-term careers in astronomy; \Nleft\ left and went into careers outside astronomy; \Npostdoc\ were still postdocs or in short-term contract-based positions at the time the analysis was conducted (late 2018).

Our study focuses on the transition in and out of the postdoctoral phase, and so we record only the \emph{first} long-term position (and not later career moves or promotions.)  However, we did also note any cases in which an individual left the field \emph{after} securing a long-term astrophysics position.  These were quite rare (12 men and 2 women, out of \Nastro\ total hires), suggesting that ``long term'' employment (as we have defined it) does indeed represent a the start of a lifetime career in the discipline.

\section{Analysis and Results}

\subsection{Career Outcomes}

In Table 1 we provide a detailed breakdown of the job and employer classifications for men and women who ultimately found stable employment (based on the nature of the first such long-term job following their PhD/postdoc, our definition of ``outcome'').  We also provide the numbers for temporary positions and the numbers of omitted individuals.

Most job and employer classes do not show any statistically significant difference in gender demographics relative to the overall fraction of women in the study (30\%).  Only for astronomers employed at small colleges are the numbers inconsistent with the overall F/M ratio to greater than $2.5\sigma$ (this role contains a higher fraction of women than expected by random chance).  The fraction of men vs.\ women who left the field overall is also the same: \mpctleave\% for men and \fpctleave\% for women\footnote{This is based on the relative numbers of individuals who were hired or left the field and does not include current postdocs.  Some of these will also be hired (or will leave) in the longer term: this will slightly change these statistics, but the hazard model (\S \ref{sec:hazard}) suggests that the change will be gender-independent and not more than a few percent.  Formally, we estimate that \pctleavetwelvem\% of men and \pctleavetwelvef\% of women leave the field within 12 years.}.

\subsection{Career Hiring Times}
\label{sec:hazard}

\begin{figure}
   \begin{minipage}{0.49\textwidth}
        \centering
        \includegraphics[width=0.95\textwidth]{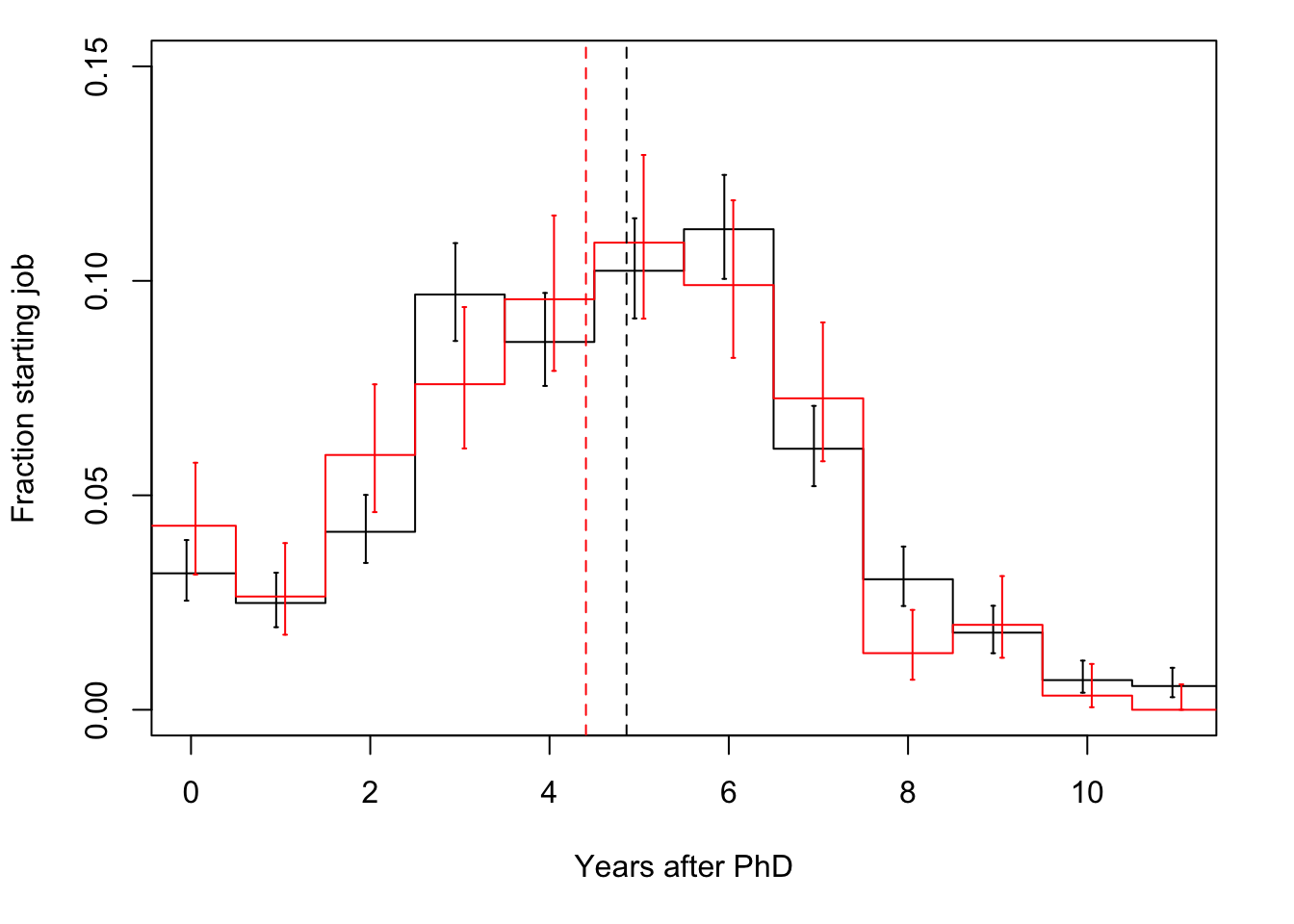}
    \end{minipage}\hfill
    \begin{minipage}{0.49\textwidth}
        \centering
        \includegraphics[width=0.95\textwidth]{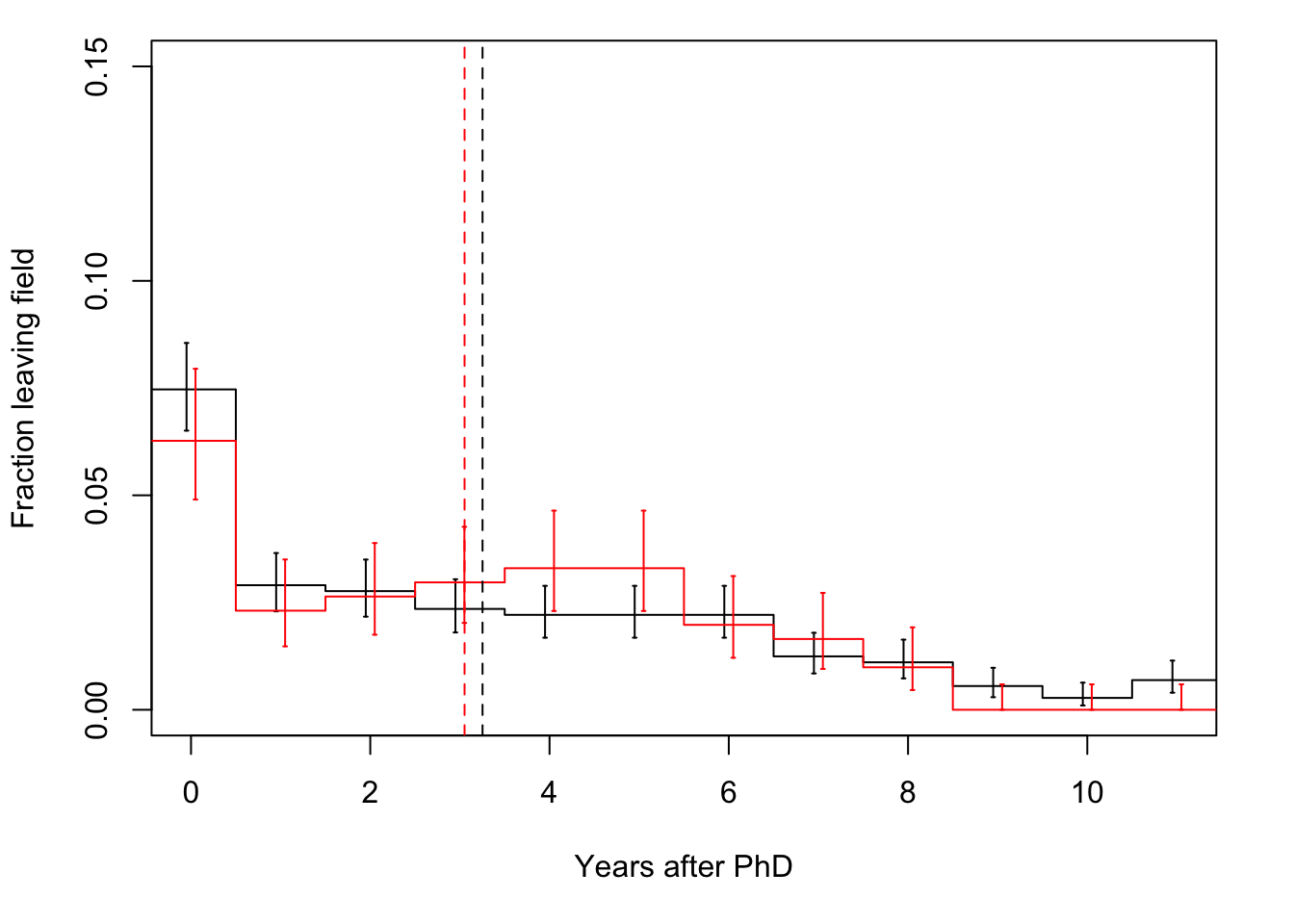}
    \end{minipage}
\center
\caption{Histograms of recorded times (years after PhD) at which PhDs either: (left) progressed from term-limited to long-term or permanent positions within astronomy, or (right) left the field to pursue other employment.  Histograms are normalized using total counts for each gender (regardless of outcome).  Error bars show 67\% binomial confidence intervals and dashed vertical lines show the means.  Male astronomers are shown in black and female astronomers in red.\label{hist}}
\end{figure}

\begin{figure}
\centering
\includegraphics[width=0.73\textwidth]{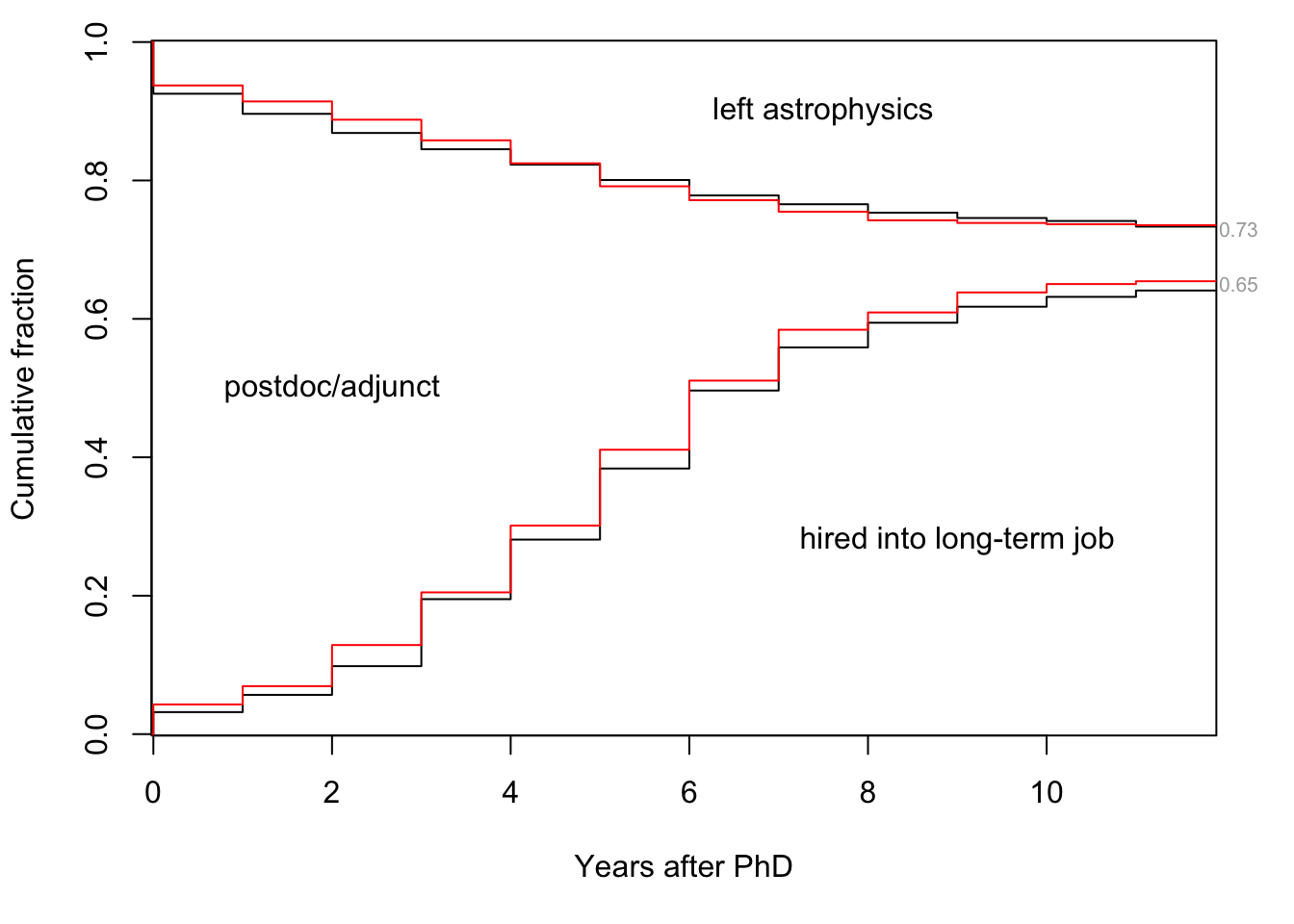}
\center
\caption{Career status by year after PhD for men (black) and women (red) astronomers.  The middle wedge indicates the fraction still in temporary positions; the top wedge indicates astronomers who left the field while the bottom wedge indicates those who secured long-term employment.  Years $>7$ have been corrected for incompleteness using our hazard model.\label{cdf}}
\end{figure}

Similar overall outcomes could nevertheless conceal gender differences in the paths to those outcomes, and the time required to achieve stable employment is of interest on its own.  To address this, we plot a pair of histograms in Figure \ref{hist}.  The histogram at left shows the times of hires \emph{into} astronomy careers; that at right shows the times at which graduates \emph{left} astronomy to pursue another career.  Hiring into astronomy shows a steady rise out to the 6th postdoctoral year and then sharply drops, with relatively few hires occurring after the 7th year.  The distribution of times at which graduates left the field shows a peak at $t=0$ years post-PhD and then a steady decline between $t=1-10$ years.  There is no obvious difference between the profiles between genders, although a formal t-test provides a marginally significant difference between the mean time to an astrophysics job between men ($t_M$ = \mmean$\pm$\mmeanunc\ yr) and women ($t_F$~=~\fmean$\pm$\fmeanunc~yr): ($\Delta t$ = \gap\ $\pm$ \gapunc\ yr; $p=\gapp$)\footnote{Throughout this paper we employ 95\% confidence intervals for quoted uncertainties, and 67\% confidence intervals for plotted error bars.}.  This is significantly less than the 1.1-year gap measured by F18 using rumor mill data ($\Delta t=1.1$ yr is ruled out at $p$ = \pFlaherty).  
An alternate representation of this data is presented in Figure \ref{cdf}, which shows the cumulative fraction of the sample in temporary positions, long-term positions, or non-astronomy positions as a function of year post-PhD.  For years $>6$ the data are incomplete (e.g., for individuals in the sample who earned their PhDs in 2012, only 6 years have elapsed, so we do not know which group they will be in after 7 or 8 years).  We use the statistics for late-year hirings/leavings based on an annualized hazard model (see below) to project the future career tracks of individuals who were still in temporary positions at the time of the study.  With or without this correction, there is no apparent difference between the two gender groups in either hiring or leaving rates.   After 12 years, \pcthiretwelve\% of astronomers have obtained long-term positions; \pctleavetwelve\% have left the field; and \pcttemptwelve\% are still in postdoc, adjunct, or short-term soft-money positions.

Time-to-hire is a form of survival data, and is dealt with most appropriately using survival analysis.  We perform two complementary forms of survival analysis, one for each potential outcome (hiring into a long-term astronomy career, or leaving astronomy).\footnote{We did not consider any covariates in either analysis, although we did investigate using the PhD year and/or the number of PhD-associated first-author publications as additional explanatory variables.  While both variables are significantly correlated with hiring time, this did not qualitatively change any of the conclusions.  For simplicity we use only the gender-only model.}  For the first analysis we model the hiring times within astronomy, taking the times at which postdocs left astronomy to be right-censored measurements (lower limits, reflecting the fact that an individual who decided to leave astronomy at year $N$ may indeed have eventually found a long-term job had they remained in the field).  For the second analysis we model the times at which graduates left the field, taking the times of hiring within astronomy to be the right-censored measurements (reflecting the possibility that had they not been hired during year $N$ and instead remained as a postdoc, they may have left the field in some future year).  Current postdocs are treated as right-censored in both cases (with a lower-limit equal to the time between the year of their PhD and 2018\footnote{For a few postdocs whose CV's were out of date and who could not confirmed to be in the same role in 2018, we used the time between the PhD year and the CV date instead.}).  Analysis was performed using the \texttt{survival} package in R.

\begin{figure}
   \begin{minipage}{0.49\textwidth}
        \centering
        \includegraphics[width=0.95\textwidth]{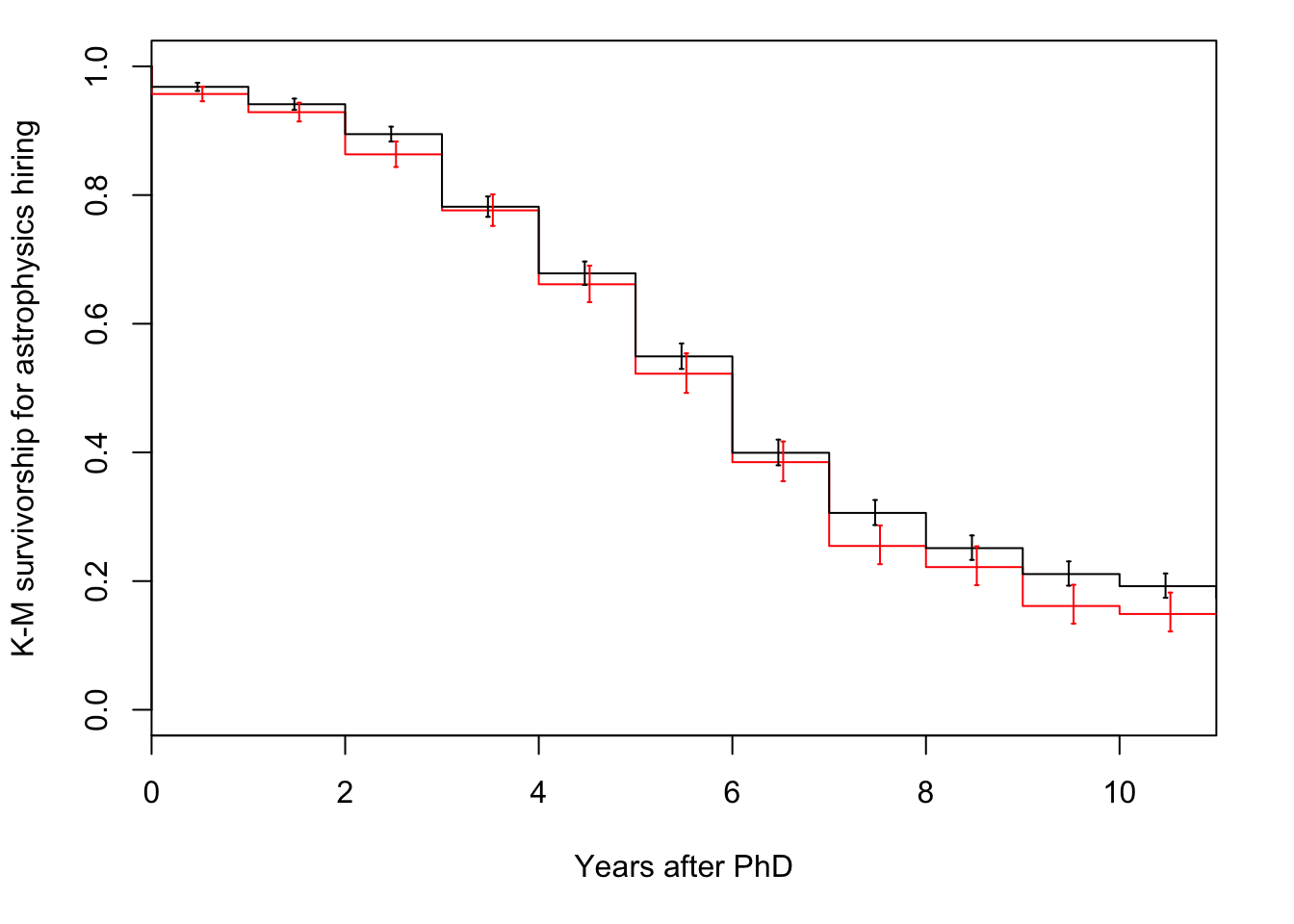}
            \end{minipage}\hfill
    \begin{minipage}{0.49\textwidth}
        \centering
        \includegraphics[width=0.95\textwidth]{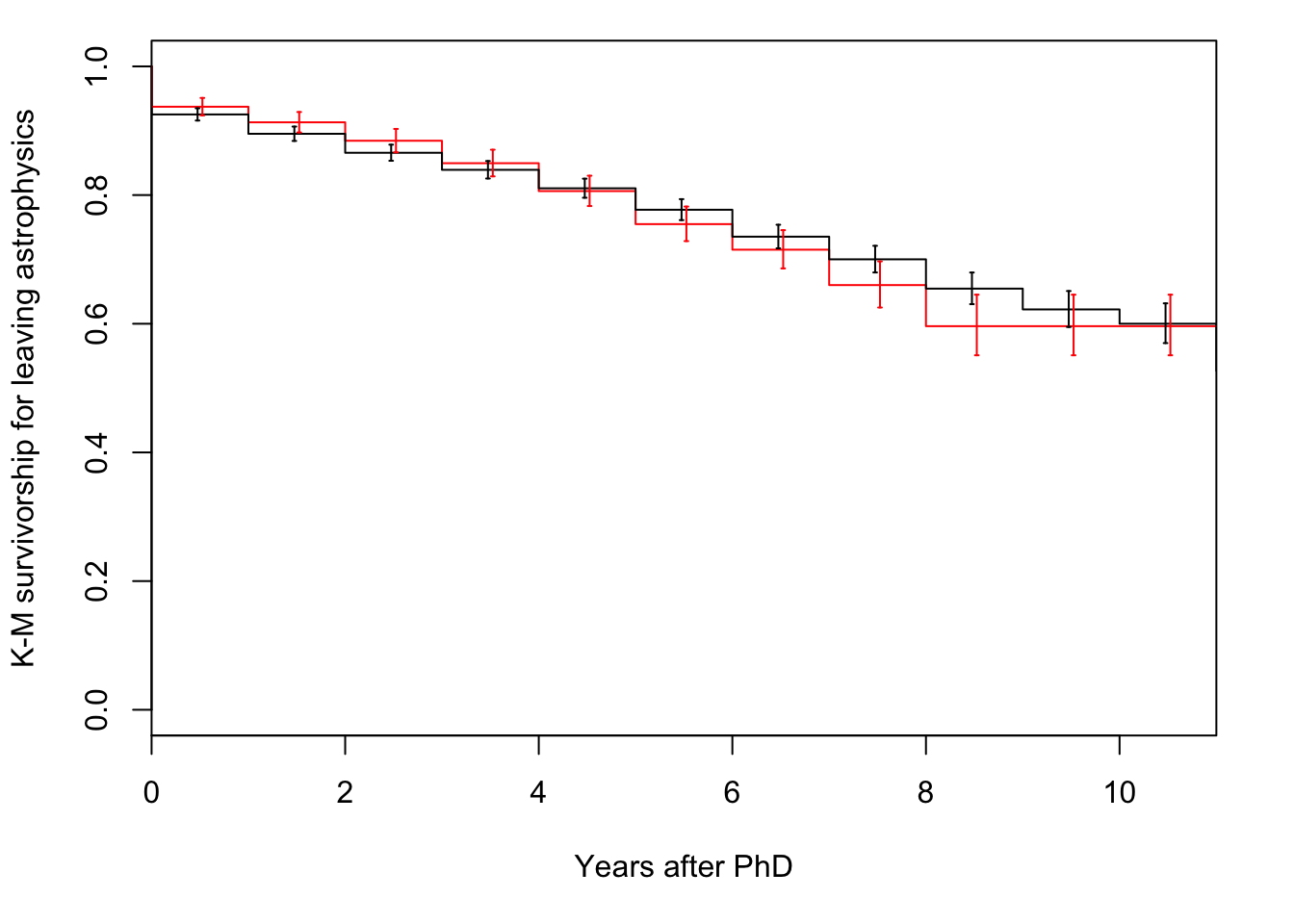}
    \end{minipage}
\center
\caption{Kaplan-Meier survivorship curves for astronomy PhDs (effectively, the cumulative distribution function corrected for incompleteness and alternative outcomes).  The left version shows the hiring time into astronomy careers; the right considers leaving the field (hiring into other careers). \label{km}}
\end{figure}

The Kaplan-Meier estimator\footnote{The Kaplan-Meier estimator \citep{KM} is a cumulative distribution corrected for censored measurements (lower limits).}
for hiring \emph{within} astronomy is plotted in Figure \ref{km} at left, and for \emph{leaving} astronomy at right.  This shows how quickly the pools of male (black) and female (red) postdocs are depleted by long-term hiring and by leaving, independent of each other: the plot at left can be thought of as the probability that a postdoc who arbitrarily refuses to ever consider any alternative career has not yet been hired by year N; the plot at right can be thought of as the probability of a postdoc who arbitrarily refuses to ever apply for long-term astronomical positions having left the field by year N.  There is no obvious gender difference between either pair of profiles.

\begin{figure}
   \begin{minipage}{0.49\textwidth}
        \centering
        \includegraphics[width=0.95\textwidth]{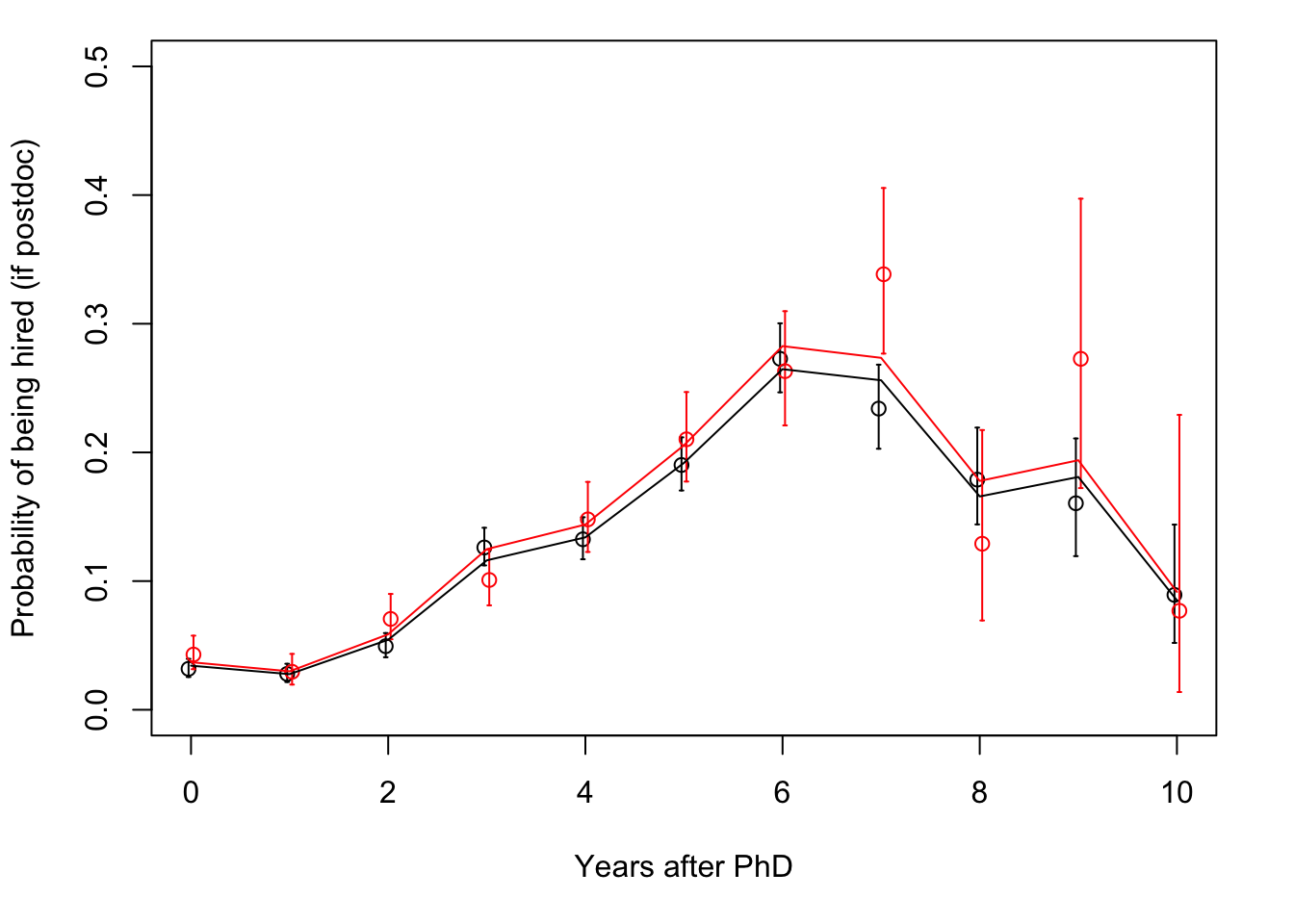}
    \end{minipage}\hfill
    \begin{minipage}{0.49\textwidth}
        \centering
        \includegraphics[width=0.95\textwidth]{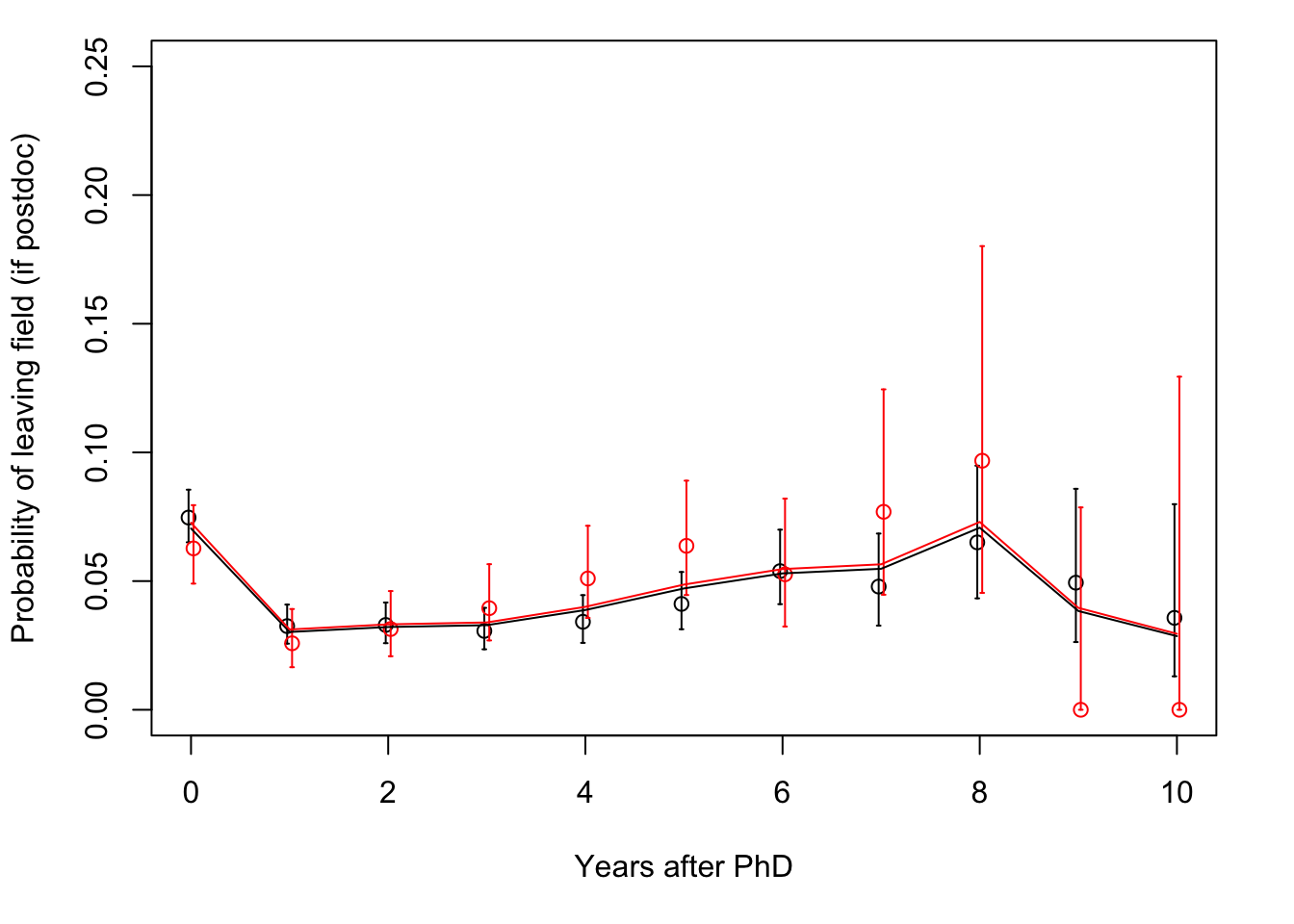}
    \end{minipage}
\center
\caption{Hazard curves for astronomy PhDs (points), along with the best-fit proportional-hazard model (lines).  This shows the probability of a postdoc/adjunct being hired into a long-term astronomy position (left) or leaving the field (right) at a given year post-PhD.  Women and men are hired and leave the field at essentially identical rates. \label{hazard}}
\end{figure}

To quantify this, we fit a Cox proportional-hazards model to the survival times for each case, treating gender as a categorical independent variable.  We confirm the lack of any significant difference: the hazard ratio for hiring (the relative probability of being hired for a female postdoc versus a male postdoc given the same year post-PhD; a ratio of 1 signifies no gender difference) is $H_{F/M} = \Hfm^{+\Hfmu}_{-\Hfml}$, while the hazard ratio for leaving (the relative probability of leaving the field for a female postdoc relative to a male postdoc of the same year post-PhD) is $L_{F/M} = \Lfm^{+\Lfmu}_{-\Lfml}$.

The corresponding annualized Cox hazard curves are presented in Figure \ref{hazard}.  This shows the probability of an astronomer \emph{who has not yet been hired already} being hired as a function of year-post-PhD (at left), or the probability of an un-hired astronomer leaving the field as a function of year-post-PhD (at right).

The hiring probability rises steadily, flattens at years 6--7, and then declines slowly (subject to the small-number statistics of very advanced postdocs).  This may come as a surprise given Figure 1, but it is expected: while the absolute number of (e.g.) 8th year postdocs hired each year is relatively few, this primarily reflects the fact that there are few 8th year postdocs to begin with: the \emph{probability} of a postdoc who has reached that stage being hired each year is comparable to a 4th year postdoc (although at 20\% it is not high in an absolute sense, and it does decline in subsequent years).  Additionally, the histograms in Figure \ref{hist} are not corrected for incompleteness/censorship that artificially depresses the counts at $>6$ years.  

The leaving probability shows a peak at $t=0$ years (corresponding to PhDs who went into industry immediately with no postdoc) and then drops to a few percent per year.  It rises gradually over the subsequent years, but always remains below 10\% per year.

We also repeated our survival analysis model for hiring into R1 tenure-track jobs specifically (treating all other forms of hiring as right-censored measurements).  We again found no significant gender difference in hiring rate ($H_{F/M} = 0.98^{+0.42}_{-0.29}$), although the constraints are weaker due to the smaller sample size.

\section{Conclusions}

In summary, there is no evidence for any significant difference in career outcomes between male and female astronomy PhDs in the United States.  The fraction of graduates pursuing postdocs, the fraction of those postdocs hired into long-term positions each year, and the fraction of those postdocs who leave the field each year, all show no gender differences.  The types of astrophysics employers show no differences either, except that women are slightly more likely than men to obtain positions at small colleges.  Quantitatively, we rule out any systematic difference between men and women in astronomy hiring rates greater than 30\% and any difference in the rate of leaving the field greater than 40\%.

Our results are consistent with the relative fractions of women reported by \cite{Hughes+2014} (i.e., that women represent approximately 30\% of PhD students, postdocs, and assistant professors) and with their indirect survival analysis of early-career advancement (Table 2 of that work).  They do not directly explain the reasons for the lower fraction of women ($\sim$15\%) in more advanced career roles.  However, we do note that the PhD numbers by gender show a large increase in the fraction of women over the period of the study (from 15\% in 2000--2001 to 34\% in 2011--2012), suggesting that a primary cause is a lower fraction of women in earlier PhD generations relative to more recent years.  An alternative explanation is attrition of women from the system \emph{after} being hired into long-term positions.  We cannot address whether this was true in earlier generations of astronomers, but the small numbers of women (and men) who departed astrophysics after obtaining a long-term job in our sample suggests that mid/late-career attrition is probably not a major factor at the present time.

Our results do not confirm the presence of large hiring time gap found by F18.  The reasons for this are not obvious, although it may originate because of their reliance on self-reported rumor mill data.\footnote{Alternatively, it is possible that women are hired earlier but defer starting their positions for longer: F18 measures the time until an offer is made, while our analysis measures the time the job actually begins.  We consider this to be rather unlikely: deferral times are rarely longer than 1 year, so nearly all women offered a job would have to defer for $>6$ months longer than the average male hiree to explain the magnitude of the difference seen in our results.}  In any case, we firmly rule out their claim (headlined in some recent news articles\footnote{https://www.nature.com/articles/d41586-018-07018-4}) that women postdocs leave the field at three times the rate of men.

We summarize our conclusions, and their implications for the state of the field, below.

\begin{itemize}

\item Most United States astronomy PhDs (\pcthiretwelve\% after 12 years) obtain long-term jobs within the field, even for smaller and lesser-known PhD programs.  The number of astronomy PhDs is not greatly in excess of the number of careers available within the field, even if most of those careers are not tenure-track faculty positions at R1 universities (see also \citealt{Dinerstein+2011}).  Calls to stem a perceived ``overproduction'' of astrophysics PhDs should be treated with skepticism.

\item Postdocs remain attractive as candidates for faculty and other long-term positions for many years after graduation.  Postdoc competitiveness increases with time up until the 6th year after PhD, and declines only slowly thereafter.   While lengthy postdocs are not uncommon, they do appear to leave candidates better equipped to compete for more secure positions within astrophysics.

\item Despite being arrayed with several sources of adversity, women perform as well as men on the astronomy job market and are not discernibly more likely to leave the field after their PhD or as postdocs.  Discrimination and other effects thus do not appear to disadvantage the career progression of junior women \emph{in aggregate} to a degree that is currently perceptible.  This may reflect the success of proactive recruitment efforts, mitigation practices, and other efforts to combat discrimination, or it may simply be a testament to the resilience of women who complete a PhD in the first place.\footnote{An alternative possibility is that women completing their PhD have superior talent on average compared to their male peers, but discrimination at later stages causes their outcomes to be similar to men.}  This also means that gender-equity efforts are not on average ``overcorrecting'' by a significant margin, since this would produce a net bias against men which we do not observe.

\item Neither the PhD-to-postdoc transition nor the postdoc-to-faculty transition represents a significant bottleneck that causes the gender skew evident in the relative numbers of male and female astronomers.  While every effort should be expended to improve the postdoctoral experience for women (as well as for men), these measures may not produce a large change in the gender demographics of professional astronomers. However, given that women who do obtain PhDs are just as likely to obtain long-term astrophysics employment as men, efforts to encourage more women to pursue and complete degrees in astronomy and physics are likely to produce a proportionate increase in the numbers of female astronomers in the long term.
\end{itemize}

Large longitudinal studies of this type in other fields and other countries will be needed to establish whether or not similar results hold in STEM disciplines outside of astrophysics, in astrophysics communities outside the United States, or within intersectional groups (e.g. ethnic and sexual-orientation minorities).  Longitudinal studies of earlier career stages (during and prior to PhD studies) are also needed, given the clear gender asymmetry in the number of graduating PhDs.  These efforts will help to shed a more general light on the impacts of gender discrimination and efforts to mitigate it.

\acknowledgements

DAP would like to thank V. Acosta, K. Alatalo, R. Beaton, B. Davies, K. Flaherty, B. Gaensler, K. T. Grasha, M. Martig, J. Michaels, and H. Ngo for their thoughts and feedback on preliminary versions of this manuscript.


\begin{thebibliography}{}
\bibitem[Bagenal(2004)]{Bagenal+2004} Bagenal, F. The Leaky Pipeline for Women in Physics and Astronomy, CSWA Status 13-19 (2004). 
\bibitem[Cech \& Blair-Loy(2019)]{Cech+2019} Cech, E.~A. \& Blair-Loy, M. 2019, PNAS, in press \url{https://doi.org/10.1073/pnas.1810862116}
\bibitem[Dinerstein(2011)]{Dinerstein+2011} Dinerstein, H.~L.\ 2011, Bulletin of the American Astronomical Society, 43, 145.02 
\bibitem[Flaherty(2018)]{F18} Flaherty, K. 2018, arXiv:1810.01511
\bibitem[Hoffman(2004)]{Hoffman+2004} Hoffman, J. L. and Urry, M.  Portrait of a Decade: Results from the 2003 CSWA Survey of Women in Astronomy, CSWA Status 1 \& 8--12 (2004)
\bibitem[Hughes(2014)]{Hughes+2014} Hughes, A.M. 2014, Status Newsletter, ed. N. Morrison 
\bibitem[Ivie(2005)]{Ivie+2005} R. Ivie and K. N. Ray, \emph{Women in physics and astronomy}, \url{http://www.aip.org/statistics/reports/women-physicsand-astronomy-2005}
\bibitem[Ivie, White, \& Chu(2016)]{Ivie+2016} Ivie, R., White, S. and Chu, R.~Y. 2016, Physical Review of Physics Education Research, 12, 020109
\bibitem[Kaplan \& Meier(1958)]{KM} Kaplan, E. L.; Meier, P. (1958).  J. Amer. Statist. Assoc., 53, 282, 457--481. \url{https://doi.org/10.2307/2281868}
\bibitem[Metcalfe et al.(2008)]{Metcalfe+2008}
Metcalfe, T. S. 2008.  PASP, 120, 229
\bibitem[Moss-Racusin et al.(2012)]{MossRacusin+2012} Moss-Racusin, C.A., et al. 2012. Proceedings of the National Academy of Sciences, 109, 41, 16474 
\bibitem[National Academies of Sciences(2018)]{NA2018} National Academies of Sciences, Engineering, and Medicine. 2018, Sexual Harassment of Women: Climate, Culture, and Consequences in Academic Sciences, Engineering, and Medicine. Washington, D.C.: The National
Academies Press. \url{https://doi.org/10.17226/24994}
\bibitem[National Research Council(2010)]{NRC2010} National Research Council, \emph{Gender Differences at Critical Transitions in the Careers of Science, Engineering, and Mathematics Faculty} (National Academies Press,
Washington, DC, 2010).
\bibitem[Shaumann(2017)]{Shaumann2017} Shauman, K. A. 2017, Social Sciences, 6, 24
\bibitem[Sleeth(2017)]{Sleeth+2017} Sleeth, K.M. 2017 {\it Results from Sexual Harassment Survey 2017}, National Postdoctoral Association, Rockville, MD \url{http://sites.nationalacademies.org/cs/groups/pgasite/documents/webpage/pga_182104.pdf}
\end{thebibliography}
\end{document}